\documentclass{cupconf}

\usepackage[dvips]{graphicx}

  \checkfont{eurm10}
  \iffontfound
    \IfFileExists{upmath.sty}
      {\typeout{^^JFound AMS Euler Roman fonts on the system,
                   using the 'upmath' package.^^J}%
       \usepackage{upmath}}
      {\typeout{^^JFound AMS Euler Roman fonts on the system, but you
                   don't seem to have the}%
       \typeout{'upmath' package installed. CUPconf.cls can take advantage
                 of these fonts,^^Jif you use 'upmath' package.^^J}%
      }
  \else
  \fi

  \checkfont{msam10}
  \iffontfound
    \IfFileExists{amssymb.sty}
      {\typeout{^^JFound AMS Symbol fonts on the system, using the
                'amssymb' package.^^J}%
       \usepackage{amssymb}%
       \let\le=\leqslant  \let\leq=\leqslant
         
      }{}
  \fi

  \IfFileExists{amsbsy.sty}
    {\typeout{^^JFound the 'amsbsy' package on the system, using it.^^J}%
     \usepackage{amsbsy}}
    {}

\newsavebox{\astrutbox}
\sbox{\astrutbox}{\rule[-5pt]{0pt}{20pt}}

  \title[Migration]{Planetary Migration}

  \author[P.J. Armitage \& W.K.M. Rice]
  {P\ls H\ls I\ls L\ls I\ls P\ns J.\ns 
  A\ls R\ls M\ls I\ls T\ls A\ls G\ls E$^{1,2}$\ns \and  
  W.\ls K.\ls M.\ns R\ls I\ls C\ls E$^{3}$\ns}

  \affiliation{$^1$JILA, 440 UCB, University of Colorado, Boulder CO 80309\\
               $^2$Department of Astrophysical and Planetary Sciences, 
               University of Colorado,\\ Boulder CO 80309\\
               $^3$Institute of Geophysics and Planetary Physics and Department 
               of Earth Sciences, University of California, Riverside CA 92521}

\begin{document}

\maketitle

\begin{abstract}
Gravitational torques between a planet and gas in the protoplanetary 
disk result in orbital migration of the planet, and modification 
of the disk surface density. Migration via this mechanism is 
likely to play an important role in the formation 
and early evolution of planetary systems. 
For masses comparable to those of observed giant extrasolar planets, 
the interaction with the 
disk is strong enough to form a gap, leading to coupled evolution 
of the planet and disk on a viscous time scale (Type~II migration). 
Both the existence of hot Jupiters, and 
the statistical distribution of observed orbital radii, are consistent 
with an important role for Type~II migration 
in the history of currently observed systems. We discuss the 
possibility of improving constraints on migration by including 
information on the host stars' metallicity, and note that migration 
could also form a population of massive planets at {\it large} 
orbital radii that may be indirectly detected via their influence 
on debris disks.
For lower mass planets with $M_p \sim M_\oplus$, surface density 
perturbations created by the planet are small, and migration in a 
laminar disk is 
driven by an intrinsic and apparently robust asymmetry between 
interior and exterior torques. Analytic and numerical 
calculations of this Type~I migration are in reasonable 
accord, and predict rapid orbital decay during the final 
stages of the formation of giant planet cores. The difficulty 
of reconciling Type~I migration with giant planet formation 
may signal basic errors in our understanding of protoplanetary 
disks, core accretion, or both. We discuss physical effects that 
might alter Type~I behavior, in particular the possibility that  
for sufficiently low masses ($M_p \rightarrow 0$)  
turbulent fluctuations in the gas surface density dominate the  
torque, leading to random walk migration of 
very low mass bodies. 
\end{abstract}

\firstsection 
\section{Introduction}
The extremely short orbital period of 51~Pegasi (\cite[Mayor \& Queloz 1995]{mayor95}) 
and the other hot Jupiters pose a problem for planet formation, 
not only because such systems bear little resemblance to the 
Solar System, but more fundamentally because the high temperatures  
expected in the protoplanetary disk at radii $a < 0.1 \ {\rm AU}$ largely 
preclude the possibility of in situ formation. Disk models by \cite{bell97} 
show that for typical T~Tauri accretion rates of 
$\dot{M} \sim 10^{-8} \ M_\odot {\rm yr}^{-1}$ 
(\cite[Gullbring et al. 1998]{gullbring98}) the midplane temperature 
interior to 0.1~AU exceeds 1000~K, destroying ices and, for the 
very closest in planets, even dust. At least the cores of these 
hot Jupiters must therefore have formed elsewhere, and subsequently 
migrated inward. Migration is also likely to 
have occurred for the larger population of 
extrasolar planets that now lie within the snow 
line in their parent disks (\cite[Bodenheimer, Hubickyj \& Lissauer 2000]{bodenheimer00}), 
though this is a more model-dependent statement since both 
the location of the snow line (\cite[Sasselov \& Lecar 2000]{sasselov00}) 
and its significance for giant planet formation remain uncertain.

Orbital migration of planets involves a loss of angular momentum 
to either gas or other solid bodies in the system. Three main 
mechanisms have been proposed, all of which involve purely 
gravitational interactions (aerodynamic  
drag, which is central to the orbital evolution of meter-scale rocks, 
is negligible for planetary masses). The first is gravitational 
interaction between the planet and the gas in the protoplanetary 
disk. This leads to angular momentum exchange between 
the planet and the gas, and resulting orbital evolution 
(\cite[Goldreich \& Tremaine 1980; Lin, Bodenheimer \& 
Richardson 1996]{goldreich80,lin96}). Since gas giant 
planets, by definition, formed at an epoch when the protoplanetary 
disk was still gas-rich, this type of migration is almost 
unavoidable. It is the main subject of this article. However, 
further migration could also occur later on, after the gas 
disk has been dissipated, as a consequence of the gravitational 
scattering of either planetesimals 
(\cite[Murray et al. 1998]{murray98}) or other massive planets 
(\cite[Rasio \& Ford 1996; Weidenschilling \& Marzari 1996; 
Lin \& Ida 1997; Papaloizou \& Terquem 2001]{rasio96,weidenschilling96,
lin97,papaloizou01}). Some orbital evolution from 
planetesimal scattering is inevitable, 
given that the formation of massive planets is highly likely 
to leave a significant mass of smaller bodies in orbits close 
enough to feel perturbations from the newly formed giant. In 
the Solar System, planetesimal scattering could have 
allowed substantial outward migration 
(\cite[Thommes, Duncan \& Levison 1999]{thommes99}) of Uranus and 
Neptune -- which have a small fraction of the Solar System's 
angular momentum -- while 
simultaneously raising the eccentricities and inclinations of 
all the giant planets to values consistent with those 
observed (\cite[Tsiganis et al. 2005]{tsiganis05}). Although 
this is an attractive theory for the architecture of the outer 
Solar System, invoking a scaled-up version of this process 
as the origin of the hot Jupiters is problematic. To drive 
large-scale migration of the typically rather massive planets 
seen in extrasolar planetary systems would require a comparable  
mass of planetesimals interior to the initial orbit of the planet. 
Such a planetesimal disk would in turn imply the prior existence of a   
rather massive gas disk, which would likely be more 
effective at causing migration than the planetesimals. Similar 
reservations apply to models of planet-planet scattering, which 
is only able to yield a population of planets at small orbital 
radii if multiple planet formation (with the planets close 
enough that they are unstable over long periods) is common. 
That said, the observation that most extrasolar planets have 
significantly eccentric orbits -- which currently defies 
explanation {\em except} as an outcome of planet-planet 
scattering (\cite[Ford, Rasio \& Yu 2003]{ford03}) -- may 
mean that at least some scattering-driven migration occurs 
in the typical system.

\begin{figure}
 \includegraphics[width=\textwidth]{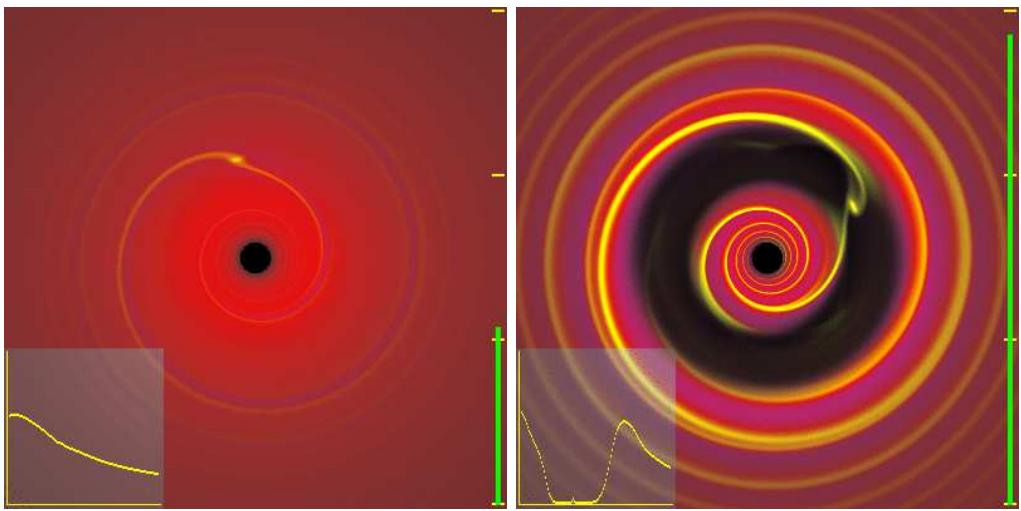}
 \vspace{0cm}
 \caption{An illustration of the interaction between a planet on a 
 fixed circular orbit with a laminar (non-turbulent) protoplanetary 
 disk, computed from a two-dimensional $(r,\phi)$ hydrodynamic 
 simulation with a locally isothermal equation of state and a 
 constant kinematic viscosity. In the left-hand panel showing 
 the regime of Type~I migration, a relatively low mass planet excites 
 a noticeable wave in the disk gas but does not significantly perturb the 
 azimuthally averaged surface density profile (shown as the inset 
 graph). In contrast, a $10 \ M_J$ planet (right-hand panel) clears 
 an annular gap in the disk, within which the surface density is 
 a small fraction of its unperturbed value. As the disk evolves 
 over a viscous time scale, the planet is predicted to track the 
 motion of the gas (either inward or outward) while remaining within 
 the gap. This is Type~II migration. For a movie showing the 
 transition between these regimes, go to {\tt http://jilawww.colorado.edu/$\sim$pja/planet{\underline{ }}migration.html}.}
 \label{armitage_f1}
\end{figure} 

Figure~\ref{armitage_f1} illustrates how a planet on a circular 
orbit interacts with the protoplanetary disk. The 
planet perturbs the gas as it passes by the planet, with angular 
momentum transport taking place at the locations of resonances 
in the disk -- radii where a characteristic disk frequency is 
related to the planet's orbital frequency. For relatively low 
mass perturbers, the interaction launches a trailing spiral 
wave in the gas disk, but is not strong enough to significantly 
perturb the azimuthally averaged surface density profile. In 
this regime, described as Type~I migration, 
angular momentum transport between the planet and the gas 
occurs while the planet remains embedded within the protoplanetary 
disk. The rate of migration is controlled by the sum of the torques 
arising from the inner and outer Lindblad and corotation resonances, 
which is generally non-zero (if the sum 
happened to be close to zero, the planet would act as a source of 
angular momentum transport in the disk 
(\cite[Goodman \& Rafikov 2001]{goodman01}) while remaining in 
place). For the parameters (sound speed, efficiency of angular momentum 
transport) that are believed to be appropriate for protoplanetary 
disks, Type~I migration occurs for planet masses $M_p \lesssim 0.1 \ M_J$, 
where $M_J$ is the mass of Jupiter (\cite[Bate et al. 2003]{bate03}), 
and is most rapid as this critical mass is approached (e.g. \cite[Ward 1997]{ward97}). 
As a result, it is likely to play a particularly important role in the final 
assembly of giant planet cores.

At higher masses -- $M_p \gtrsim 0.1 \ M_J$ -- the angular momentum removal / 
deposition at the planet's inner / outer Lindblad resonances is strong 
enough to repel gas from an annular region surrounding the planet's orbit, 
forming a gap in which the surface density is reduced compared to its 
unperturbed value. For planets of a Jupiter mass and above, the gap is 
almost entirely evacuated (e.g. the right-hand panel of Figure~\ref{armitage_f1}), 
although mass may continue to flow in a stream across the gap to enter 
the planet's Roche lobe (\cite[Artymowicz \& Lubow 1996; Lubow, Seibert \& 
Artymowicz 1999]{artymowicz96,lubow99}). The location of the inner and outer 
edges of the gap are set by a balance between angular momentum exchange with 
the planet (which tends to widen the gap), and internal stresses within 
the protoplanetary disk (`viscosity', which tends to close it). Under 
most circumstances, this balance acts to lock the planet into the 
long term viscous evolution of the disk gas. At radii where the disk 
gas is moving inward the planet migrates toward the star, all the 
while remaining within its gap (\cite[Lin \& Papaloizou 1986]{lin86}). 
This is Type~II migration, which differs from Type~I not only in the 
presence of a gap, but also because the rate depends directly on the 
efficiency of angular momentum transport within the protoplanetary 
disk. 

In addition to these well-established migration regimes, qualitatively 
different behavior may occur at very low masses, and at masses 
intermediate between the Type~I and Type~II regimes. At sufficiently 
low masses (probably of $\sim 10 \ M_\oplus$ and below) the persistent 
Type~I torque may be overwhelmed by {\em random} torques from surface 
density perturbations in a turbulent protoplanetary disk 
(\cite[Nelson \& Papaloizou 2004; Laughlin, Steinacker \& Adams 2004]
{nelson04,laughlin04}). This process, which is similar to the 
heating of Galactic stars by 
transient spiral arms (\cite[Carlberg \& Sellwood 1985]{carlberg85}), 
may lead to rapid random walk migration on top of Type~I drift. 
This could have important consequences for both 
core accretion (\cite[Rice \& Armitage 2003]{rice03}) and 
terrestrial planet formation. Another uncertain regime lies 
at the transition mass ($M_p \approx 0.1 \ M_J$) between 
Type~I and Type~II migration, where a partial gap exists and 
corotation torques can be highly significant. \cite{masset03} 
suggested that the corotation torques could drive an instability 
in the {\em direction} of migration, which if confirmed would 
be extremely important for massive planet formation. Subsequent 
higher resolution simulations by \cite{dangelo05} have 
demonstrated that extremely high resolution is needed in 
order to accurately capture torques arising from within and 
around the Hill sphere. For a Saturn mass planet, the resolution 
had to be increased to approximately 50 grid zones per Hill radius 
before a numerically converged solution for the migration rate 
was attained. Lower resolutions yielded artificially rapid 
migration. \cite{dangelo05} conclude from this that although 
torques from the near-planet region are significant, they do 
not appear to dramatically increase the migration rate.

\section{Type~I migration}
In the Type~I regime the perturbation induced by the 
planet in the gas disk remains small, and the net torque  
has surprisingly little dependence on the microphysics of the 
protoplanetary disk (\cite[e.g. Goldreich \& Tremaine 1978; 
Lin \& Papaloizou 1979]{goldreich78,lin79}). In particular, 
viscosity -- normally the most uncertain element of a 
protoplanetary disk model -- enters only indirectly via 
its influence on the magnitude and radial gradient of 
the surface density and sound speed. Generically, the 
net torque scales with the planet mass as $T \propto M_p^2$, 
so that the migration time scale at a given radius scales 
as $\tau \propto M_p^{-1}$. Type~I migration therefore 
becomes increasingly important as the planet mass increases, 
and is fastest just prior to gap opening (the onset of 
which {\em does} depend on the disk viscosity). Despite 
its attractive lack of dependence on uncertain disk 
physics, the actual calculation of the net torque is 
technically demanding, and substantial improvements 
have been made only recently. Here, we summarize a 
few key results -- the reader is directed to 
the original papers (primarily by Artymowicz, Ward, and 
their collaborators) for full details of the calculations.

\subsection{Analytic calculations}
The simplest calculation of the Type~I torque 
(\cite[Goldreich \& Tremaine 1978, 1979, 1980]{goldreich78,
goldreich79,goldreich80}) neglects 
significant pressure effects in the disk close to the 
planet, and is therefore valid for low $m$ resonances. In this 
approximation, Lindblad resonances occur at radii 
in the disk where the 
epicyclic frequency $\kappa$ is an integral multiple $m$ 
of the angular velocity in a frame rotating with the 
planet at angular velocity $\Omega_p$. For a Keplerian 
disk, this condition,
$$
 D(r) \equiv \kappa^2 - m^2 \left( \Omega - \Omega_p \right)^2 = 0,
$$ 
can be simplified using the fact that $\kappa = \Omega$. The 
resonances lie at radii,
$$
 r_L = \left( 1 \pm {1 \over m} \right)^{2/3} r_p
$$
where $r_p$ is the planet's orbital radius. The lowest order 
resonances lie at $r = 1.587 r_p$ and $r = 0.630 r_p$, 
but an increasingly dense array of high $m$ resonances 
lie closer to the planet. Resonances at $r < r_p$ add 
angular momentum to the planet, while those at $r > r_p$ 
remove angular momentum. The torque at each resonance $T_m$ can be 
evaluated in terms of a {\em forcing function} $\Psi_m$ as,
$$
 T_m = -\pi^2 m \Sigma { \Psi_m^2 \over {r {\rm d}D / {\rm d}r} },
$$ 
where $\Sigma$ is the gas surface density. 
Explicit expressions for $\Psi_m$ are given by \cite{goldreich79}. 
The net torque is obtained by evaluating the torque at each resonance, 
and then summing over all $m$. 

For Type~I migration the behavior of gas close to the planet, where 
$\Psi_m$ is largest, is critical. Accurately treating this regime 
requires elaborations of the basic \cite{goldreich79} approach. 
For a razor thin two-dimensional 
disk model, - the effects of radial pressure and density 
gradients, the calculation is described in \cite[Ward (1997), and 
references therein (especially Ward 1988; Artymowicz 1993a, 199b; Korycansky \& 
Pollack 1993)]{ward97,ward88,artymowicz93a,artymowicz93b,korycansky93}. 
These papers include the shifts in the location of Lindblad resonances 
due to both radial and azimuthal pressure gradients, which become 
significant effects at high $m$. For Lindblad resonances, the 
result is that the dominant torque arises from wavenumbers 
$m \simeq r_p / h$, where $h$, the vertical disk scale height, is 
given in terms of the local sound speed $c_s$ by $h = c_s / \Omega_p$. 
The fractional net torque $2 \vert T_{\rm inner} + T_{\rm outer} \vert / 
(\vert T_{\rm inner} \vert + \vert T_{\rm outer} \vert)$ can
be as large as 50\% (\cite[Ward 1997]{ward97}), 
with the outer resonances dominating and driving rapid inward 
migration. Moreover, the small shifts in the locations of resonances 
that occur in disks with different radial surface density profiles 
conspire so that the net torque is only weakly dependent on the 
surface density profile. This means that the predicted rapid inward migration 
occurs for essentially any disk model in which the sound speed 
decreases with increasing radius (\cite[Ward 1997]{ward97}). 
Corotation torques -- which vanish in the oft-considered disk 
models with $\Sigma \propto r^{-3/2}$ -- can alter the magnitude 
of the torque but are not sufficient to reverse the sign 
(\cite[Korycansky \& Pollack 1993; Ward 1997]{korycansky93,ward97}). 

The observation that the dominant contribution to the total torque 
comes from gas that is only $\Delta r \simeq h$ away from the planet 
immediately implies that a two-dimensional representation of the 
disk is inadequate, even for protoplanetary disks which are 
geometrically thin by the usual definition ($h/r < 0.1$, so that 
pressure gradients, which scale as order $(h/r)^2$, are sub-percent 
level effects). Several new physical effects come into play in a 
three-dimensional disk:
\begin{itemize}
\item[1.]
The perturbing potential has to be averaged over the vertical 
thickness of the disk, effectively reducing its strength for 
high $m$ resonances (\cite[Miyoshi et al. 1999]{miyoshi99}). 
\item[2.]
The variation of the scale height with radius decouples the 
radial profile of the midplane density from that of the 
surface density.
\item[3.]
Wave propagation in three-dimensional disks is fundamentally 
different from that in two dimensions, if the vertical 
structure of the disk departs from isothermality 
(\cite[Lubow \& Ogilvie 1998; Bate et al. 2002]{lubow98,bate02}).
\end{itemize}
\cite{tanaka02} have computed the interaction between a planet 
and a three-dimensional isothermal disk, including the first 
two of the above effects. Both Lindblad and corotation torques 
were evaluated. They find that the net torque is reduced by a 
factor of 2-3 as compared to a corresponding two-dimensional 
model, but that migration remains inward and is typically 
rapid. Specifically, defining a local migration time scale 
via,
$$
 \tau = {r_p \over {-{\dot r}_p}}, 
$$ 
\cite{tanaka02} find that for a disk in which $\Sigma \propto r^{-\beta}$, 
$$
 \tau = (2.7 + 1.1 \beta)^{-1} {M_* \over M_p} { M_* \over 
 {\Sigma r_p^2} } \left( {c_s \over {r_p \Omega_p} } \right)^2 
 \Omega_p^{-1}.
$$
As expected, the time scale is inversely proportional to the 
planet mass and the local surface density. Since the bracket 
is $\sim (h/r_p)^2$, the time scale also decreases quite rapidly 
for thinner disks, reflecting the fact that the peak torque 
arises from closer to the planet as the sound speed drops. 

Although still limited by the assumption of isothermality, 
the above expression represents the current `standard' estimate of the 
Type~I migration rate. Although slower than two-dimensional 
estimates, it is still rapid enough to pose a potential 
problem for planet formation via core accretion. There is 
therefore interest in studying additional physical effects 
that might reduce the rate further. The influence of disk 
turbulence is discussed more fully in the next Section; here 
we list some of the other effects that might play a role:
\begin{itemize}
\item[1.]
{\bf Realistic disk structure models}. The run of density and 
temperature in the midplane of the protoplanetary disk is not 
a smooth power-law, due to sharp changes in opacity and, potentially, 
the efficiency of angular momentum transport (\cite[Gammie 1996]{gammie96}). 
\cite{menou04} have calculated Type~I rates in Shakura-Sunyaev type 
disk models, and find that even using the standard Lindblad torque 
formula there exist regions of the disk where the migration rate is locally 
slow. Such zones could be preferred sites of planet formation.
\item[2.]
{\bf Thermal effects}. \cite{jangcondell05} find that the dominant 
non-axisymmetric thermal effect arises from changes to the stellar 
illumination of the disk surface in the vicinity of the planet. This 
effect is most important at large radii, and can increase the migration 
time scale by up to a factor of 2 at distances of a few AU.
\item[3.]
{\bf Wave reflection}. The standard analysis assumes that waves 
propagate away from the planet, and are dissipated before they 
reach boundaries or discontinuities in disk properties that might 
reflect them back toward the planet. \cite{tanaka02} observe that 
reflection off boundaries has the potential to substantially 
reduce the migration rate. We note, however, that relaxation 
of vertical isothermality will probably lead to wave dissipation 
in the disk atmosphere within a limited radial distance 
(\cite[Lubow \& Ogilvie 1998]{lubow98}), and thereby reduce the 
possibility for reflection.
\item[4.]
{\bf Accretion}. Growth of a planet during Type~I migration is 
accompanied by a non-resonant torque, which has been evaluated 
by \cite{nelson03a}. If mass is able to accrete freely on to 
the planet, \cite{bate03} find from three-dimensional 
simulations that $\dot{M}_p \propto M_p$ for $M_p \lesssim 10 \ M_\odot$, 
with a mass doubling time that is extremely short (less than 
$10^3$~yr). In reality, it seems likely the planet will be 
unable to accept mass at such a rapid rate, so the mass 
accretion rate and resulting torque will then depend on the 
planet structure.
\item[5.]
{\bf Magnetic fields}. The dominant field component in magnetohydrodynamic 
disk turbulence initiated by the magnetorotational instability is 
toroidal (\cite[Balbus \& Hawley 1998]{balbus98}). \cite{terquem03} 
finds that gradients in plausible toroidal magnetic fields can 
significantly alter the Type~I rate, and in some circumstances 
(when the field decreases rapidly with $r$) stop migration. More 
generally, a patchy and variable toroidal field might lead to rapid variations 
in the migration rate. Whether this, or density fluctuations induced 
by turbulence, is the primary influence of disk fields is unclear.
\item[6.]
{\bf Multiple planets}. The interaction between multiple planets 
has not been studied in detail. \cite{thommes05} notes that 
low mass planets, which would ordinarily suffer rapid Type~I 
migration, can become captured into resonance with more 
massive bodies that are themselves stabilized against rapid decay 
by a gap. This may be important for understanding multiple 
planet formation (and we have already noted that there is 
circumstantial evidence that multiple massive planet formation 
may be common), though it does not explain how the {\em first} 
planet to form can avoid rapid Type~I inspiral.
\item[7.]
{\bf Disk Eccentricity}. Type~I migration in an axisymmetric 
disk is likely to damp planetary eccentricity. However, 
it remains possible that the protoplanetary disk itself might 
be spontaneously unstable to development of eccentricity 
(\cite[Ogilvie 2001]{ogilvie01}). \cite{papaloizou02} has 
shown that Type~I migration can be qualitatively altered, 
and even reversed, if the background flow is eccentric.
\end{itemize}

\subsection{Numerical simulations}
Hydrodynamic simulations of the Type~I regime within a 
shearing sheet geometry have been presented by \cite{miyoshi99}, 
and in cylindrical geometry by \cite{dangelo02},  
\cite{dangelo03}, and \cite{nelson03b}, with the latter 
paper focusing on the transition between Type~I and 
Type~II behavior. The most comprehensive work to date 
is probably that of \cite{bate03}, who simulated in 
three dimensions the interaction with the disk of planets in the 
mass range $1 \ M_\oplus \leq M_p \leq 1 \ M_J$. 
The disk model had $h/r = 0.05$, 
a Shakura-Sunyaev (1973) $\alpha$ parameter 
$\alpha = 4 \times 10^{-3}$, and a fixed locally isothermal 
equation of state (i.e. $c_s = c_s(r)$ only). This setup 
is closely comparable to that assumed in the calculations 
of \cite{tanaka02}, and very good agreement was obtained 
between the simulation results and  
the analytic migration time scale. Based on this, it seems reasonable to 
conclude that {\em within the known limitations imposed 
by the restricted range of included physics}, current 
calculations of the Type~I rate are technically reliable. 
Given this, it is interesting to explore the 
consequences of rapid Type~I migration for 
planet formation itself.

\subsection{Consequences for planet formation}
The inverse scaling of the Type~I migration time scale with planet mass 
means that the most dramatic effects for planet formation 
occur during the growth of giant plant cores via core accretion 
(\cite[Mizuno 1980]{mizuno80}). In the baseline calculation of 
\cite{pollack96}, which does not incorporate migration, runaway 
accretion of Jupiter's envelope is catalyzed by the slow formation 
of a $\approx 20 \ M_\oplus$ core over a period of almost 10~Myr. 
This is in conflict with models by \cite{guillot04}, which show that 
although Saturn has a core of around 15~$M_\oplus$, Jupiter's core 
is observationally limited to at most $\approx 10 \ M_\odot$, leading to discussion 
at this meeting of several ways to reduce the theoretically predicted 
core mass. Irrespective of the uncertainties, however, it seems 
inevitable that planets forming via core accretion pass through a 
relatively slow stage in which the growing planet has a mass 
of $5-10 \ M_\oplus$. This stage is vulnerable to Type~I drift.

\begin{figure}
 \includegraphics[width=\textwidth]{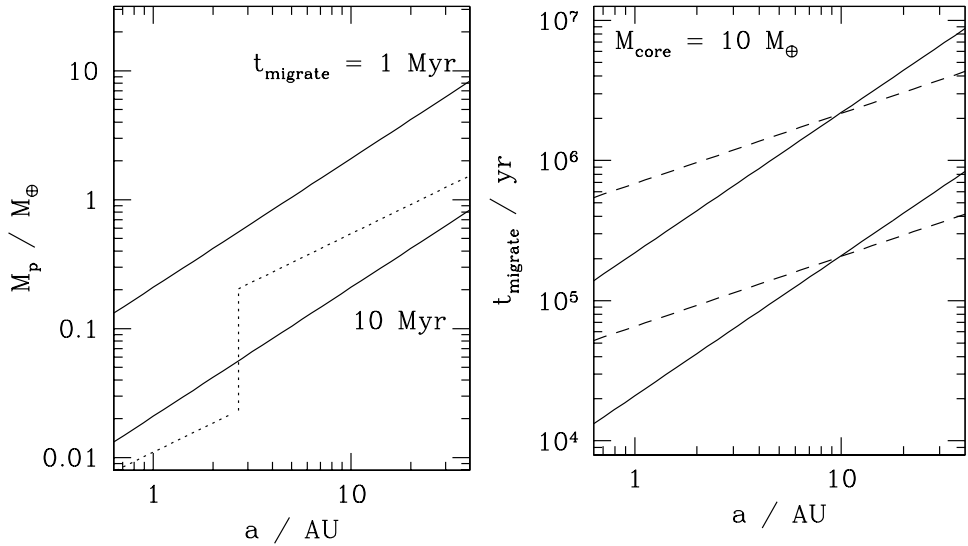}
 \vspace{-6cm}
 \caption{Left hand panel: planet mass for which the Type~I migration 
 time scale at different radii equals 1~Myr (upper solid curve) and 
 10~Myr (lower curve). The gas disk is assumed to have a surface density 
 profile $\Sigma \propto r^{-3/2}$, and a mass within 30~AU of 0.01~$M_\odot$.
 The dotted line shows an estimate of the isolation mass in the same disk 
 model, assuming Solar metallicity and a snow line at 2.7~AU. Right hand panel:
 the migration time scale for a 10~$M_\oplus$ core in protoplanetary disks 
 with surface density profiles of $\Sigma \propto r^{-3/2}$ (solid lines) 
 and $\Sigma \propto r^{-1}$ (dashed lines). For each model, the lower  
 curve shows the time scale for a disk with a gas mass of 0.01~$M_\odot$ 
 within 30~AU, while the upper curve shows results for an absolute 
 minimum mass gas disk with only 1~$M_J$ within 30~AU.}
 \label{armitage_f2}
\end{figure}

Figure~\ref{armitage_f2} shows the migration time scale for a 
$10 \ M_\oplus$ planet within gas disks with surface 
density profiles of $\Sigma \propto r^{-1}$ (very roughly that 
suggested by theoretical models, \cite[e.g. Bell et al. 1997]{bell97}) 
and $\Sigma \propto r^{-3/2}$ (the minimum mass Solar Nebula 
profile of \cite[Weidenschilling 1977]{weidenschilling77}). We 
consider disks with integrated gas masses (out to 30~AU) of 
$0.01 \ M_\odot$ and 1~$M_J$. The latter evidently represents 
the absolute minimum gas mass required to build Jupiter or a 
typical extrasolar giant planet. The torque formula of 
\cite{tanaka02} is used to calculate the migration time 
scale $\tau$. As is obvious from the Figure, migration from 5~AU on 
a time scale of 1~Myr -- significantly less than either the typical 
disk lifetime (\cite[Haisch, Lada \& Lada 2001]{haisch01}) or 
the duration of the slow phase of core accretion -- is 
inevitable for a core of mass 10~$M_\oplus$, even if there 
is only a trace of gas remaining at the time when the envelope  
is accreted. For more reasonable gas masses, the typical migration 
time scale at radii of a few AU is of the order of $10^5$~yr.
Another representation of this is to note that in the giant 
planet forming region, we 
predict significant migration ($\tau = 10~{\rm Myr}$) for 
masses $M_p \gtrsim 0.1 \ M_\oplus$, and rapid migration ($\tau = 1~{\rm Myr}$) 
for $M_p \gtrsim 1 \ M_\oplus$. We can also plot, for the same 
disk model, an estimate of the isolation mass (\cite[Lissauer 
1993; using the gas to planetesimal surface density scaling 
of Ida \& Lin 2004a]{lissauer93,ida04}). The isolation mass 
represents the mass a growing planet can attain by 
consuming only those planetesimals within its feeding zone -- as 
such it is reached relatively rapidly in planet formation 
models. Outside the snow line, the migration time scale for a 
planet (or growing core) at the isolation mass is typically 
a few Myr, reinforcing the conclusion that migration is 
inevitable in the early stage of giant planet formation. By 
contrast, in the terrestrial planet region, interior to the 
snow line, planets need to grow significantly beyond isolation 
before rapid migration ensues. It is therefore possible for 
the early stages of terrestrial planet formation to occur in 
the presence of gas with only limited Type~I migration, while 
the final assembly of terrestrial planets happens subsequently 
in a gas poor environment.

Does Type~I migration help or hinder the growth of giant 
planets via core accretion? This question remains open, 
despite a history of investigations stretching back at 
least as far as papers by \cite{hourigan84} and 
\cite{ward89}. Two competing effects are at work:
\begin{itemize}
\item[1.]
A migrating core continually moves into planetesimal-rich regions 
of the disk that have not been depleted by the core's prior growth. 
This depletion is, in part, responsible for the slow growth of 
Jupiter in static core calculations. Calculations suggest that a 
rapidly migrating core can capture of the order of 10\% of the 
planetesimals it encounters (\cite[Tanaka \& Ida 1999]{tanaka99}), 
with the collision fraction increasing with migration velocity. 
Slow migration velocities allow for inward shepherding of the 
planetesimals rather than capture, and a low accretion rate 
(\cite[Ward \& Hahn 1995; Tanaka \& Ida 1999]{ward95,tanaka99}). 
\item[2.]
A migrating core must reach the critical core mass before 
it is lost to the star, on a time scale that, as we noted 
above, can be an order of magnitude or more smaller than 
the gas disk lifetime. 
Unfortunately, a high accretion rate of planetesimals  
increases the critical core mass needed before runaway gas 
accretion starts and, at fixed accretion rate, the critical mass 
also increases as the core moves inward (\cite[Papaloizou \& 
Terquem 1999]{papaloizou99}). Migration therefore favors a 
high rate of planetesimal accretion, but often hinders attaining 
the critical core mass needed for envelope accretion.
\end{itemize}

Calculations of giant planet formation including steady core migration 
have been presented by several groups, including \cite{papaloizou99}, 
\cite{papaloizou00}, \cite{alibert04} and \cite{alibert05}. The 
results suggest that, for a single growing core, Type~I migration 
at the standard rate of \cite{tanaka02} is simply too fast to 
allow giant planet formation to occur across a reasonable range 
of radii in the protoplanetary disk. Most cores are lost to the 
star or, if they manage to accrete envelopes at all, do so at 
such small radii that their ultimate survival is doubtful. More 
leisurely migration, on the other hand, at a rate suppressed 
from the \cite{tanaka02} value by a factor of 10 to 100, {\em helps} 
core accretion by mitigating the depletion effect that acts 
as a bottleneck for a static core (\cite[Alibert et al. 2005]{alibert05}). 

This difficulty in reconciling our best estimates of the Type~I 
migration rate with core accretion signals that something is 
probably wrong with one or both of these theories. Three 
possibilities suggest themselves. First, the Type~I migration 
rate may be a substantial overestimate, by an order of magnitude 
or more. If so, there is no need for substantial changes to 
core accretion theory or to protoplanetary disk models. We have 
already enumerated a list of candidate physical reasons for why 
the Type~I rate may be wrong, though achieving a sufficiently 
large suppression does not seem to be straightforward. Second, 
angular momentum transport may be strongly suppressed in the 
giant planet formation region by the low ionization fraction, 
which suppresses MHD instabilities that rely on coupling 
between the gas and the magnetic field (\cite[Gammie 1996; 
Sano et al. 2000]{gammie96,sano00}). An almost inviscid 
disk could lower the threshold for gap opening 
sufficiently far that the slow stage of core accretion 
occurred in a gas-poor environment (\cite[elements of 
such a model have been explored by Rafikov 2002; Matsumara \& 
Pudritz 2005]{rafikov02,matsumara05}). Finally, and perhaps 
most attractively, it may be possible to find a variant of 
core accretion that is compatible with undiluted Type~I 
migration. For a single core, we have studied simple models 
in which random walk migration leads to large fluctuations 
in the planetesimal accretion rate and an early onset 
of criticality (\cite[Rice \& Armitage 2003]{rice03}). In the 
more realistic situation where  
multiple cores are present in the disk, it is possible that the 
early loss of the first cores to form (at small radii just 
outside the snow line) could evacuate the inner disk 
of planetesimals, allowing subsequent cores to reach 
their critical mass and accrete envelopes as they migrate 
inward. Further work is needed to explore such scenarios 
quantitatively.

\section{Stochastic migration in turbulent disks}
To a first approximation the efficiency of angular momentum transport 
(unless it is very low) has little impact on the predicted Type~I 
migration rate. This assumes, however, that the disk is laminar. 
More realistically, angular momentum transport itself derives 
from turbulence, which is accompanied by a spatially and temporally  
varying pattern of density fluctuations in the protoplanetary disk. These 
fluctuations will exert {\em random} torques on planets of any 
mass embedded within the disk, in much the same way as transient 
spiral features in the Galactic disk act to increase the velocity 
dispersion of stellar populations (\cite[Carlberg \& Sellwood 1985]{carlberg85}). 
If we assume that the random torques 
are uncorrelated with the presence of a planet, then the random 
torques' linear scaling with planet mass will dominate over 
the usual Type~I torque (scaling as $M_p^2$) for sufficiently 
low masses. The turbulence will then act to increase the 
velocity dispersion of collisionless bodies, or, in the presence 
of damping, to drive a random walk in the semi-major axis 
of low mass planets.

\begin{figure}
 \includegraphics[width=\textwidth]{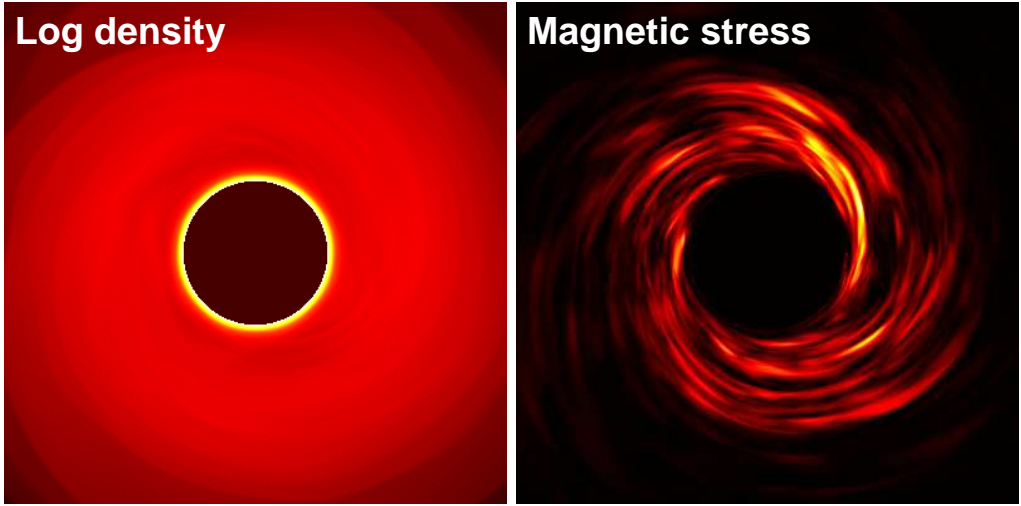}
 \vspace{0cm}
 \caption{Structure in a turbulent disk accretion flow, from a 
 global cylindrical MHD simulation (\cite[after Armitage 1998]{armitage98}). 
 The flow is here visualized using density (left hand panel) and the vertically averaged Maxwell 
 stress $\langle B_r B_\phi \rangle$ (right hand panel) as tracers. In magnetically 
 active disks, transient density and magnetic field fluctuations are 
 present across a wide range of spatial scales.}
 \label{armitage_f3}
\end{figure}

To go beyond such generalities, and in particular to estimate the 
crossover mass between stochastic and Type~I migration, we need to 
specify the source of turbulence in the protoplanetary disk. MHD 
disk turbulence (see Figure~\ref{armitage_f3} for an illustration) 
driven by the magnetorotational instability 
(\cite[Balbus \& Hawley 1998]{balbus98}) provides a well-understood 
source of outward angular momentum transport in sufficiently 
well-ionized disks, and has been used as a model system for 
studying stochastic migration by \cite{nelson04} and by 
\cite{laughlin04}. Density fluctuations in MHD disk turbulence 
have a typical coherence time of the order of the orbital period, 
and as a consequence are able to exchange angular momentum 
with an embedded planet across a range of disk radii (not 
only at narrow resonances). The study by \cite{nelson04} was based 
on both global ideal MHD disk simulations, with an aspect ratio of 
$h/r = 0.07$, and local shearing box calculations. The global 
runs realized an effective Shakura-Sunyaev $\alpha = 7 \times 10^{-3}$, 
which if replicated in a real disk would be consistent with 
observational measures of T~Tauri disk evolution 
(\cite[Hartmann et al. 1998]{hartmann98}). For all masses 
considered in the range $3 \ M_\oplus \le M_p \le 30 \ M_\oplus$, 
the {\em instantaneous} torque on the planet from the MHD 
turbulent disk exhibited large fluctuations in both magnitude 
and sign. Averaging over $\approx 20$ orbital periods, the 
mean torque showed signs of converging to the Type~I rate, 
although the rate of convergence was slow, especially for 
the lowest mass planets in the global runs. These properties are 
generally in accord with other studies of the variability 
of MHD disk turbulence (\cite[Hawley 2001; Winters, Balbus \& Hawley 2003a]
{hawley01,winters03}). Very roughly, the \cite{nelson04} simulations 
suggest that up to $M_p \sim 10 \ M_\oplus$ the random walk component 
dominates steady Type~I drift over time scales that substantially 
exceed the orbital period. 

We caution that existing studies of the stochastic migration regime 
are unrealistic. Ideal MHD is not a good approximation for the 
protoplanetary disk at those radii where planet formation occurs, 
and there may be dead zones in which MHD turbulence and angular momentum 
transport is highly suppressed (\cite[Gammie 1996; Sano et al. 2000; 
Fromang, Terquem \& Balbus 2002; Salmeron \& Wardle 2005]{gammie96,
sano00,fromang02,salmeron05}). We also note that a significant 
random migration component, if it indeed adds to rather than 
supplanting steady Type~I migration, does nothing (on average) to 
help the survival prospects of low mass planets.
Nevertheless, if turbulent fluctuations 
(whatever their origin) do occur in the disk, the resulting random 
walk migration could be important for planet formation. 
In the terrestrial planet region, stochastic migration might deplete 
low mass planetary embryos that would be relatively immune to 
ordinary Type~I migration, while simultaneously promoting radial 
mixing and collision of planetesimals. For giant planet formation, 
a significant random component to core migration would have the 
effect of creating large fluctuations in the planetesimal 
accretion rate, while also potentially acting to diffuse the 
planetesimal surface density. Large fluctuations in the 
planetesimal accretion rate favor the early onset of rapid gas 
accretion, and allow for the final core mass to be substantially 
smaller than would be expected in the case of a static core 
(\cite[Rice \& Armitage 2003]{rice03}).

\section{Type~II migration}
\subsection{Conditions for the onset of Type~II migration}
In a viscous disk, the threshold between Type~I and Type~II 
migration can be derived by equating the time scale for 
Type~I torques to open a gap (in the absence of viscosity) with 
the time scale for viscous diffusion to fill it in 
(\cite[Goldreich \& Tremaine 1980; Papaloizou \& Lin 1984]
{goldreich80,papaloizou84}). For a gap of width $\Delta r$ 
around a planet with mass ratio $q=M_p/M_*$, orbiting at 
distance $r_p$, Type~I torques can open the gap on a 
timescale (\cite[Takeuchi, Miyama \& Lin 1996]{takeuchi96})
$$
 \tau_{\rm open} \sim {1 \over {m^2 q^2} } 
 \left( { {\Delta r} \over r_p } \right)^2 \Omega_p^{-1}.
$$
Viscous diffusion will close the gap on a time scale,
$$
 \tau_{\rm close} \sim { {\Delta r}^2 \over \nu},
$$
where $\nu$, the kinematic viscosity,is usually written 
as $\nu = \alpha c_s^2 / \Omega_p$ (\cite[Shakura \& Sunyaev 1973]
{shakura73}). Equating $\tau_{\rm open}$ and $\tau_{\rm close}$, 
and noting that waves with $m \approx r_p / h$ dominate the 
Type~I torque, the condition for gap opening becomes,
$$
 q \gtrsim \left( { h \over r } \right)_p^2 \alpha^{1/2}.
$$  
For $h/r \simeq 0.05$ and $\alpha = 10^{-2}$, the transition 
(which simulations show is not very sharp) occurs 
at $q_{\rm crit} \sim 2.5 \times 10^{-4}$, i.e. 
close to a Saturn mass for a Solar mass star. Since the 
most rapid Type~I migration occurs when $q \approx q_{\rm crit}$, 
we can also estimate a {\em minimum} migration time scale by 
combining the above expression with the time scale formula of 
\cite{tanaka02}. This yields,
$$
 \tau_{\rm min} \sim (2.7 + 1.1 \beta)^{-1} { M_* \over 
 {\Sigma r_p^2} } \alpha^{-1/2}  
 \Omega_p^{-1}, 
$$
and is almost independent of disk properties other than the 
local mass. 

\subsection{The rate of Type~II migration}
Once a planet has becomes massive enough to open a gap, orbital 
evolution is predicted to occur on the same local time scale as the 
protoplanetary disk. The radial velocity of gas in the disk is,
$$
 v_r = - { {\dot{M} \over {2 \pi r \Sigma} } }, 
$$
which for a steady disk away from the boundaries can be written  
as,
$$
 v_r = - { 3 \over 2 } {\nu \over r}.
$$
If the planet enforces a rigid tidal barrier at the outer edge 
of the gap, then evolution of the disk will force the orbit 
to shrink at a rate $\dot{r}_p \simeq v_r$, provided that the 
local disk mass exceeds the planet mass, i.e. $\pi r_p^2 \Sigma \gtrsim M_p$. 
This implies a nominal Type~II migration time scale, valid for 
{\em disk dominated migration} only,
$$
 \tau_0 = {2 \over {3 \alpha} } \left( { h \over r } \right)_p^{-2} 
 \Omega_P^{-1}.
$$ 
For $h/r = 0.05$ and $\alpha = 10^{-2}$, the migration time scale 
at 5~AU is of the order of 0.5~Myr.

In practice, the assumption that the local disk mass exceeds 
that of the planet often fails. For example, a $\beta=1$ disk with a mass of 
$0.01 \ M_\odot$ within 30~AU has a surface density profile,
$$
 \Sigma = 470 \left( {r \over {1 \ {\rm AU}} } \right)^{-1} \ 
 {\rm g cm}^{-2}.
$$ 
The condition that $\pi r_p^2 \Sigma = M_p$ gives an estimate of the 
radius within which disk domination ceases of,
$$
 r = 6 \left( {M_p \over M_J} \right) \ {\rm AU}.
$$ 
Interior to this radius, the planet acts as a slowly moving 
barrier which impedes the inflow of disk gas. Gas piles up behind 
the barrier -- increasing the torque -- but this process does 
not continue without limit because the interaction also deposits 
angular momentum into the disk, causing it to expand 
(\cite[Pringle 1991]{pringle91}). The end result is to 
slow migration compared to the nominal rate quoted above. 
For a disk in which the surface density can be written as 
a power-law in accretion rate and radius,
$$
 \Sigma \propto {\dot{M}}^a r^b ,
$$
\cite{syer95} define a measure of the degree of disk 
dominance,
$$
 B \equiv { {4 \pi r_p^2 \Sigma} \over M_p }.
$$ 
Then for $B < 1$ (the planet 
dominated case appropriate to small radii) the actual 
Type~II migration rate is (\cite[Syer \& Clarke 1995]{syer95}),
$$
 \tau_{II} = \tau_0 B^{-a / (1+a)} .
$$
Note that with this definition of $B$, disk dominance 
extends inward a factor of a few further than would be 
predicted based on the simple estimate given above.

Evaluating how the 
surface density depends upon the accretion rate -- and thereby determining 
the $a$ which enters into the suppression term -- requires 
a full model for the protoplanetary disk (not just knowledge of the 
slope of the steady-state surface density profile). For the 
disk models of \cite{bell97}, we find that $a \simeq 0.5$ 
at 1~AU for $\dot{M} \sim 10^{-8} \ M_\odot {\rm yr}^{-1}$. At 
this radius the model with $\alpha = 10^{-2}$ has a surface 
density of about 200~gcm$^{-2}$. For a Jupiter mass planet 
we then have $B =0.3$ and $\tau_{II} = 1.5 \tau_0$. This 
is a modest suppression of the Type~II rate, but the effect 
becomes larger at smaller radii (or for more massive planets). 
It slows the inward drift of massive planets, and allows 
a greater chance for them to be stranded at sub-AU 
scales as the gas disk is dissipated.

These estimates of the Type~II migration velocity assume 
that once a gap has been opened, the planet maintains an 
impermeable tidal barrier to gas inflow. In fact, simulations 
show that planets are able to accrete gas via tidal streams 
that bridge the gap (\cite[Lubow, Seibert \& Artymowicz 1999]{lubow99}). 
The effect is particularly pronounced for planets only just 
massive enough to open a gap in the first place. If the 
accreted gas does not have the same specific angular momentum 
as the planet, this constitutes an additional accretion 
torque in addition to the resonant torque. It is likely 
to reduce the Type~II migration rate further.

\subsection{Simulations}
Simulations of gap opening and Type~II migration have been 
presented by a large number of authors, with recent examples including \cite{bryden99}, \cite{lubow99}, 
\cite{nelson00}, \cite{kley01}, \cite{papaloizou01b}, \cite{dangelo02}, 
\cite{dangelo03}, \cite{dangelo03b}, \cite{bate03}, \cite{schafer04} and \cite{lufkin04}. 
These authors all assumed, for simplicity, that angular momentum transport in the 
protoplanetary disk could be represented using a microscopic viscosity. Only a 
few recent simulations, by \cite{winters03b}, \cite{nelson04} and \cite{papaloizou04},  
have directly simulated the interaction of a planet with a turbulent disk. For 
planet masses significantly above the gap opening threshold, simulations 
support the general scenario outlined above, while also finding:
\begin{itemize}
\item[1.]
{\bf Significant mass accretion} across the gap. For planet masses close 
to the gap opening threshold, accretion is surprisingly efficient, 
with tidal streams delivering gas at a rate that is comparable 
to the {\em disk} accretion rate in the absence of a planet 
(\cite[Lubow, Seibert \& Artymowicz 1999]{lubow99}). It is also 
observed that accretion cuts off rapidly as the planet mass 
increases toward $10 \ M_J$, giving additional physical motivation 
to the standard dividing line between massive planets and brown dwarfs.
\item[2.]
{\bf Damping of eccentricity}. \cite{goldreich80} noted that the 
evolution of the eccentricity of a planet embedded within a disk 
depends upon a balance between Lindblad torques, which tend to 
excite eccentricity, and corotation torques, which damp it. Recent 
analytic work (\cite[Ogilvie \& Lubow 2003; Goldreich \& Sari 2003; and 
references therein]{ogilvie03,goldreich03}) has emphasized that if the 
corotation resonances are even partially saturated, the overall balance 
tips to eccentricity excitation. To date, however, numerical simulations 
(\cite[e.g. those by Papaloizou, Nelson \& Masset 2001]{papaloizou01b}) 
exhibit damping for planetary mass bodies, while eccentricity 
growth is obtained for masses $M_p \gtrsim 20 \ M_J$ -- in the 
brown dwarf regime. In a recent numerical study, however, \cite{masset04} 
present evidence that the resolution attained by \cite{papaloizou01b} 
was probably inadequate to determine the sign of eccentricity 
evolution for Jupiter mass planets, which remains uncertain.  
Further numerical work is needed. 
\end{itemize}

\subsection{Comparison with statistics of extrasolar planetary systems}

\begin{figure}
 \includegraphics[width=\textwidth]{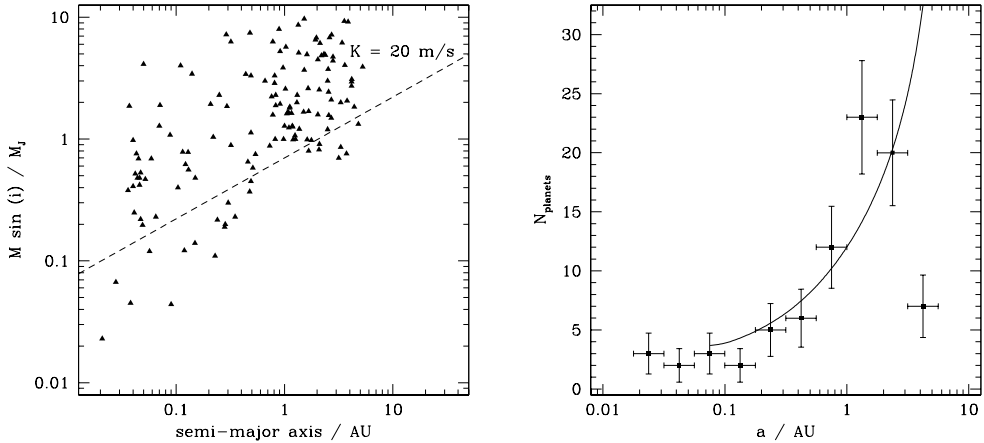}
 \vspace{-1cm}
 \caption{Left hand panel: the distribution of extrasolar planets discovered 
 via radial velocity surveys in the $a$-$M_p \sin(i)$ plane. The dashed 
 diagonal traces a line of equal ease of detectability -- planets on 
 circular orbits lying along lines parallel to this cause the same amplitude 
 of stellar radial velocity variations. Right hand panel: the number of 
 detected planets with $M_p \sin(i) > 1 \ M_J$ is plotted as a function 
 of semi-major axis. The solid curve shows the predicted distribution 
 from a pure migration model by \cite{armitage02}. The theoretical 
 curve is unaltered from the 2002 version except for an arbitrary 
 normalization.}
 \label{armitage_f4}
\end{figure}

The estimated time scale for migration of a giant planet from 5~AU 
to the hot Jupiter region is of the order of a Myr. This time scale 
is short enough -- compared to the lifetime of typical protoplanetary 
disks -- to make migration a plausible origin for hot Jupiters, while 
not being so short as to make large-scale migration inevitable (the 
latter would raise obvious concerns as to why there is no evidence 
for substantial migration of Jupiter itself). Having passed this 
initial test, it is then of interest to try and compare {\em quantitative}  
predictions of migration, most obviously the expected distributions of 
planets in mass and orbital radius, with observations. Models that 
attempt this include those by \cite{trilling98}, \cite{armitage02}, 
\cite{trilling02}, \cite{ida04} and \cite{ida04b}. Accurate 
knowledge of the biases and selection function of the radial 
velocity surveys is essential if such exercises are to be 
meaningful, making analyses such as those of \cite{cumming99} 
and \cite{marcy05} extremely valuable.

Figure~\ref{armitage_f4} shows how the distribution of known 
extrasolar planets with semi-major axis compares to the pure 
migration model of \cite{armitage02}. In this model, we 
assumed that giant planets form and gain most of their 
mass at an orbital radius (specifically 5~AU) beyond where 
most extrasolar planets are currently being detected. Once 
formed, planets migrate inward via Type~II migration and are 
either (a) swallowed by the star, or (b) left stranded at 
some intermediate radius by the dispersal of the protoplanetary 
disk. We assumed that disk dispersal happens as a consequence 
of photoevaporation (\cite[Johnstone, Hollenbach \& Bally 1998]{johnstone98}), 
and that the probability of planet formation per unit time is 
constant over the (relatively short) window during which a massive 
planet can form at 5~AU and survive without being consumed by 
the star. Although clearly oversimplified, it is interesting that 
this model continues 
to reproduce the orbital distribution of known planets out to 
radii of a few~AU, once selection effects have been taken into 
account (here, by simply ignoring low mass planets that are 
detectable only at small orbital radii). Moreover, it 
predicts that substantial {\em outward} migration ought to 
occur in disks where strong mass loss prompts an outwardly 
directed radial velocity in the giant planet forming region 
(\cite[Veras \& Armitage 2004]{veras04}). Planets at these large radii are 
potentially detectable today via their effect on debris disks (\cite[Kuchner \& 
Holman 2003]{kuchner03}).

\begin{figure}
 \includegraphics[width=\textwidth]{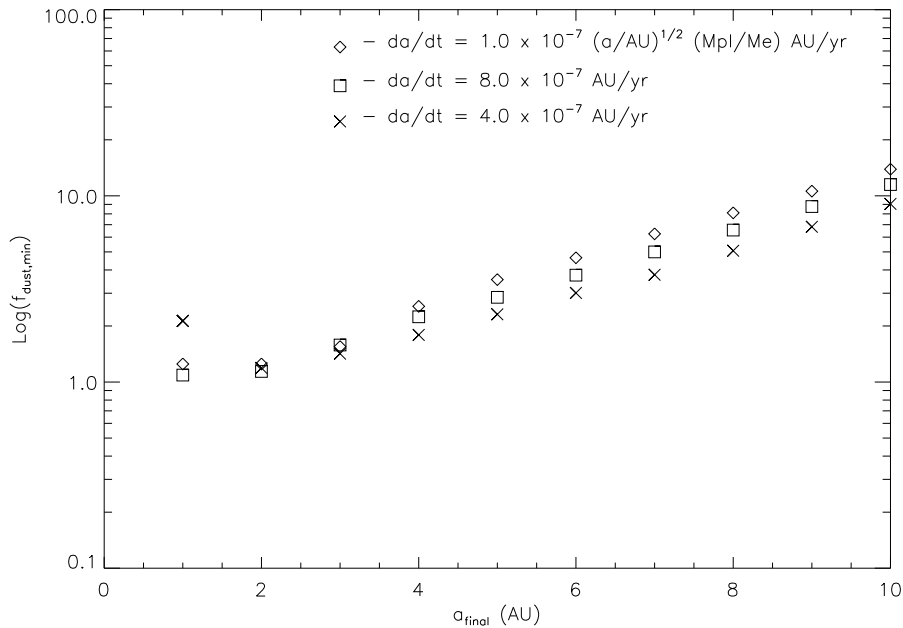}
 \vspace{-9.5cm}
 \caption{The minimum metallicity required to form a gas giant planet as 
 a function of orbital radius, from simplified core accretion calculations 
 by \cite{rice05}. The host metallicity is expressed via the parameter 
 $f_{\rm dust}$, which is proportional to $10^{\rm [Fe/H]}$. The models assume 
 that dispersion in disk metallicity dominates over dispersion 
 in either disk gas mass or disk lifetime in controlling the probability 
 of planet formation. Type~I migration of the core prior to accretion 
 of the envelope is included, using several different prescriptions 
 indicated by the different symbols, though none of the Type~I rates are as 
 large as the baseline rate of \cite{tanaka02}.}
 \label{armitage_f5}
\end{figure}

Additional clues to the role of migration in forming the 
observed population of extrasolar planets may be possible by 
combining large planet samples with knowledge of the host 
stars' metallicity. It is now clear that the frequency of 
detected planets increases rapidly with host metallicity, 
and that the measured metallicity reflects primarily the 
primordial composition of the star-forming gas rather than 
subsequent pollution of the convective envelope 
(\cite[Santos, Israelian \& Mayor 2004; Fischer \& Valenti 2005; and 
references therein]{santos04,fischer05}). The existence of 
such a correlation is not in itself surprising, given that the 
time scale for core accretion decreases quickly with increasing 
surface density of planetesimals. However, the sharpness of the 
rise in planet frequency over a fairly narrow range of stellar 
[Fe/H] {\em is} striking, since it suggests that metallicity, 
rather than variations in initial gas disk mass or gas disk lifetime, may 
well be the single most important parameter determining the 
probability of giant planet formation around a particular star.

Motivated by these observations, we have investigated the use of 
the critical or threshold metallicity for giant planet formation as 
a discriminant of different planet formation models 
(\cite[Rice \& Armitage 2005]{rice05}). Using simplified core 
accretion models similar to those of \cite{ida04}, we have 
calculated the radial dependence of the threshold metallicity 
under the assumption that disks around different stars have 
similar initial gas masses and lifetimes (this could be 
replaced with the much weaker assumption that the gas mass and 
disk lifetime are not correlated with the metallicity). A sample 
result is shown in Figure~\ref{armitage_f5}. By 
definition, planets that just manage to form as the gas disk is 
being lost suffer little or no Type~II migration, so delineating the 
threshold metallicity curve observationally can yield information 
on planet formation that is independent of Type~II migration uncertainties. 
We find that the most important influence on the shape of the threshold 
metallicity curve is probably {\em Type~I} core 
migration prior to accretion of the gas envelope. When this 
is included we derive  
a monotonically rising minimum metallicity curve beyond about 
2~AU. In the absence of significant core migration, the threshold 
metallicity is flat beyond the snow line (with a weak dependence 
on the surface density profile of planetesimals), and the location 
of the snow line may be preserved in the observed distribution 
of planets in the orbital radius / metallicity plane.

\section{Summary}
Migration via gravitational interactions between planets and 
the gaseous protoplanetary disk appears to be central to 
understanding both the formation of gas giant planets, and 
their early orbital evolution to yield the extrasolar planetary 
systems currently being observed. Although there are 
uncertainties in our understanding of migration, 
there has also been enough progress to convince us 
that Type~I migration is probably a vital  
ingredient in the formation of gas giant planets via core 
accretion. This is trivially true using the current best 
estimates of the Type~I migration time scale, but it would 
still remain important even if the rate was suppressed by 
as much as two orders of magnitude. On a similarly firm 
footing is the assertion that gas disk 
migration -- probably in the Type~II regime -- is responsible 
for the existence of most of the hot Jupiters. Although 
other migration mechanisms certainly exist, it requires 
moderate care to {\em avoid} substantial orbital evolution 
once a planet has formed in a gas disk.

Equally interesting are the major unknowns. Does turbulence 
within the disk lead to random walk migration of low mass 
bodies, and if so, is this important for terrestrial planet 
formation and / or core accretion? Do corotation torques 
qualitatively change the behavior of migrating planets 
with masses just above the gap-opening threshold? Is the 
eccentricity of massive planets excited by 
the interaction with the gas disk? A positive answer 
to any of these questions would require substantial changes 
to our overall picture of planet formation. Addressing 
them will probably require, in part, high resolution 
simulations that include more of the complex physics 
of angular momentum transport within the protoplanetary 
disk. 

\begin{acknowledgments}
This work was supported by NASA under grants NAG5-13207 and NNG04GL01G 
from the Origins of Solar Systems and Astrophysics Theory Programs, and 
by the NSF under grant AST~0407040. PJA acknowledges the hospitality of 
the Kavli Institute for Theoretical Physics, supported in part
by the NSF under grant PHY99-07949.
\end{acknowledgments}


\begin{thebibliography}{}

\bibitem[Alibert, Mordasini \& Benz (2004)]{alibert04}	
 \textsc{Alibert, Y., Mordasini, C. \& Benz, W.} 2004, 
 \emph{A\&A}, 417, L25

\bibitem[Alibert et al. (2005)]{alibert05}
 \textsc{Alibert, Y., Mordasini, C., Benz, W. \& Winisdoerffer, C.} 2005, 
 \emph{A\&A}, 434, 343
 
\bibitem[Armitage (1998)]{armitage98}
 \textsc{Armitage, P. J.} 1998, 
 \emph{ApJ}, 501, L189 
 
\bibitem[Armitage et al. (2002)]{armitage02} 	
 \textsc{Armitage, P. J., Livio, M., Lubow, S. H. \& Pringle, J. E.} 2002, 
 \emph{MNRAS}, 334, 248

\bibitem[Artymowicz (1993a)]{artymowicz93a}
 \textsc{Artymowicz, P.} 1993a, 
 \emph{ApJ}, 419, 155
 
\bibitem[Artymowicz (1993b)]{artymowicz93b}
 \textsc{Artymowicz, P.} 1993b, 
 \emph{ApJ}, 419, 166 

\bibitem[Artymowicz \& Lubow (1996)]{artymowicz96}
 \textsc{Artymowicz, P. \& Lubow, S. H.} 1996, 
 \emph{ApJ}, 467, L77
 
\bibitem[Balbus \& Hawley (1998)]{balbus98} 	
 \textsc{Balbus, S. A. \& Hawley, J. F.} 1999, 
 \emph{Reviews of Modern Physics}, 70, 1

\bibitem[Bate et al. (2003)]{bate03}	
 \textsc{Bate, M. R., Lubow, S. H., Ogilvie, G. I. \& Miller, K. A.} 2003, 
 \emph{MNRAS}, 341, 213

\bibitem[Bate et al. (2002)]{bate02} 
 \textsc{Bate, M. R., Ogilvie, G. I., Lubow, S. H. \& Pringle, J. E.} 2002, 
 \emph{MNRAS}, 332, 575
 	
\bibitem[Bell et al. (1997)]{bell97}
 \textsc{Bell, K. R., Cassen, P. M., Klahr, H. H. \& Henning, Th.} 1997,
 \emph{ApJ}, 486, 372

\bibitem[Bodenheimer, Hubickyj \& Lissauer (2000)]{bodenheimer00}
 \textsc{Bodenheimera, P., Hubickyj, O. \& Lissauer, J. J.} 2000, 
 \emph{Icarus}, 143, 2

\bibitem[Bryden et al. (1999)]{bryden99} 
 \textsc{Bryden, G., Chen, X., Lin, D. N. C., Nelson, R. P. \& Papaloizou, J. C. B.} 1999, 
 \emph{ApJ}, 514, 344
 
\bibitem[Carlberg \& Sellwood (1985)]{carlberg85} 
 \textsc{Carlberg, R. G. \& Sellwood, J. A.} 1985, 
 \emph{ApJ}, 292, 79
 
\bibitem[Cumming, Marcy \& Butler (1999)]{cumming99} 	
 \textsc{Cumming, A., Marcy, G. W. \& Butler, R. P.} 1999, 
 \emph{ApJ}, 526, 890
 
\bibitem[D'Angelo, Bate \& Lubow (2005)]{dangelo05} 	
 \textsc{D'Angelo, G., Bate, M. R. \& Lubow, S. H.} 2005, 
 \emph{MNRAS}, 358, 316
 
\bibitem[D'Angelo, Henning \& Kley (2002)]{dangelo02} 
 \textsc{D'Angelo, G., Henning, Th. \& Kley, W.} 2002, 
 \emph{A\&A}, 385, 647
 
\bibitem[D'Angelo, Henning \& Kley (2003)]{dangelo03b} 
 \textsc{D'Angelo, G., Henning, Th. \& Kley, W.} 2003, 
 \emph{ApJ}, 599, 548
 
\bibitem[D'Angelo, Kley \& Henning (2003)]{dangelo03} 
 \textsc{D'Angelo, G., Kley, W. \& Henning, Th.} 2003, 
 \emph{ApJ}, 586, 540 

\bibitem[Fischer \& Valenti (2005)]{fischer05} 
 \textsc{Fischer, D. A. \& Valenti, J.} 2005, 
 \emph{ApJ}, 622, 1102 
 
\bibitem[Ford, Rasio \& Yu 2003]{ford03} 	
 \textsc{Ford, E. B., Rasio, F. A. \& Yu, K.} 2003, 
 in {\emph Scientific Frontiers in Research on Extrasolar Planets}, 
 ASP Conference Series 294 (San Francisco), eds Drake Deming \& Sara Seager, p.~181
 
\bibitem[Fromang, Terquem \& Balbus (2002)]{fromang02}
 \textsc{Fromang, S., Terquem, C. \& Balbus, S. A.} 2002, 
 \emph{MNRAS}, 329, 18 
 
\bibitem[Gammie (1996)]{gammie96}
 \textsc{Gammie, C. F.} 1996, 
 \emph{ApJ}, 547, 355 
 
\bibitem[Goldreich \& Sari (2003)]{goldreich03} 	
 \textsc{Goldreich, P. \& Sari, R.} 2003, 
 \emph{ApJ}, 585, 1024

\bibitem[Goldreich \& Tremaine (1978)]{goldreich78} 	
 \textsc{Goldreich, P. \& Tremaine, S.} 1978, 
 \emph{Icarus}, 34, 240
 
\bibitem[Goldreich \& Tremaine (1979)]{goldreich79} 	
 \textsc{Goldreich, P. \& Tremaine, S.} 1979, 
 \emph{ApJ}, 233, 857 
 
\bibitem[Goldreich \& Tremaine (1980)]{goldreich80} 	
 \textsc{Goldreich, P. \& Tremaine, S.} 1980, 
 \emph{ApJ}, 241, 425
 
\bibitem[Goodman \& Rafikov (2001)]{goodman01} 	
 \textsc{Goodman, J. \& Rafikov, R. R.} 2001, 
 \emph{ApJ}, 552, 793
 
\bibitem[Guillot (2004)]{guillot04}	
 \textsc{Guillot, T.} 2004,
 \emph{Annual Review of Earth and Planetary Sciences}, 33, 493
 
\bibitem[Gullbring et al. (1998)]{gullbring98}
 \textsc{Gullbring, E., Hartmann, L., Briceno, C. \& Calvet, N.} 1998, 
 \emph{ApJ}, 492, 323
 
\bibitem[Haisch, Lada \& Lada (2001)]{haisch01} 
 \textsc{Haisch, K. E., Lada, E. A. \& Lada, C. J.} 2001, 
 \emph{ApJ}, 533, L153
 
\bibitem[Hartmann et al. (1998)]{hartmann98}
 \textsc{Hartmann, L., Calvet, N., Gullbring, E. \& D'Alessio, P.} 1988, 
 \emph{ApJ}, 495, 385
 
\bibitem[Hawley (2001)]{hawley01} 
 \textsc{Hawley, J. F.} 2001, 
 \emph{ApJ}, 554, 534 

\bibitem[Hourigan \& Ward (1984)]{hourigan84} 
 \textsc{Hourigan, K. \& Ward, W. R.} 1984, 
 \emph{Icarus}, 60, 29 
 
\bibitem[Ida \& Lin (2004a)]{ida04}
 \textsc{Ida, S. \& Lin, D. N. C.} 2004a,
 \emph{ApJ}, 604, 388 
 
\bibitem[Ida \& Lin (2004b)]{ida04b}
 \textsc{Ida, S. \& Lin, D. N. C.} 2004b,
 \emph{ApJ}, 616, 567  
 
\bibitem[Jang-Condell \& Sasselov (2005)]{jangcondell05}	
 \textsc{Jang-Condell, H. \& Sasselov, D. D.} 2005,
 \emph{ApJ}, 619, 1123
 
\bibitem[Johnstone, Hollenbach \& Bally (1998)]{johnstone98} 	
 \textsc{Johnstone, D., Hollenbach, D. \& Bally, J.} 1998, 
 \emph{ApJ}, 499, 758

\bibitem[Kley, D'Angelo \& Henning (2001)]{kley01} 
 \textsc{Kley, W., D'Angelo, G. \& Henning, T.} 2001, 
 \emph{ApJ}, 547, 457
 
\bibitem[Korycansky \& Pollack (1993)]{korycansky93}
 \textsc{Korycansky, D. G. \& Pollack, J. B.} 1993, 
 \emph{Icarus}, 102, 150
 
\bibitem[Kuchner \& Holman (2003)]{kuchner03} 
 \textsc{Kuchner, M. J. \& Holman, M. J.} 2003, 
 \emph{ApJ}, 588, 1110
 
\bibitem[Laughlin, Steinacker \& Adams (2004)]{laughlin04}	
 \textsc{Laughlin, G., Steinacker, A. \& Adams, F. C.} 2004, 
 \emph{ApJ}, 608, 489 
 
\bibitem[Lin, Bodenheimer \& Richardson (1996)]{lin96} 	
 \textsc{Lin, D. N. C., Bodenheimer, P. \& Richardson, D. C.} 1996, 
 \emph{Nature}, 380, 606

\bibitem[Lin \& Ida (1997)]{lin97} 
 \textsc{Lin, D. N. C. \& Ida, S.} 1997,
 \emph{ApJ}, 477, 781 
 
\bibitem[Lin \& Papaloizou (1979)]{lin79} 
 \textsc{Lin, D. N. C. \& Papaloizou, J.} 1979, 
 \emph{MNRAS}, 186, 799 
 
\bibitem[Lin \& Papaloizou (1986)]{lin86} 
 \textsc{Lin, D. N. C. \& Papaloizou, J.} 1986, 
 \emph{ApJ}, 309, 846
 
\bibitem[Lissauer (1993)]{lissauer93}
 \textsc{Lissauer, J. J.} 1993, 
 \emph{ARA\&A}, 31, 129

\bibitem[Lubow \& Ogilvie (1998)]{lubow98} 
 \textsc{Lubow, S. H. \& Ogilvie, G. I.} 1998,
 \emph{ApJ}, 504, 983

\bibitem[Lubow, Seibert \& Artymowicz (1999)]{lubow99} 
 \textsc{Lubow, S. H., Seibert, M. \& Artymowicz, P.} 1999, 
 \emph{ApJ}, 526, 1999
 
\bibitem[Lufkin et al. (2004)]{lufkin04} 
 \textsc{Lufkin, G., Quinn, T., Wadsley, J., Stadel, J. \& Governato, F.} 2004, 
 \emph{MNRAS}, 347, 421
 
\bibitem[Marcy et al. (2005)]{marcy05}
 \textsc{Marcy, G., Butler, R. P., Fischer, D. A., Vogt, S. S., Wright, J. T., 
 Tinney, C. G. \& Jones, H. R. A.} 2005, Progress of Theoretical Physics Supplement, 158, 24

\bibitem[Masset \& Ogilvie (2004)]{masset04} 
 \textsc{Masset, F. S. \& Ogilvie, G. I.} 2004, 
 \emph{ApJ}, 615, 1000
 
\bibitem[Masset \& Papaloizou (2003)]{masset03}	
 \textsc{Masset, F. S. \& Papaloizou, J. C. B.} 2003, 
 \emph{ApJ}, 588, 494 

\bibitem[Matsumara \& Pudritz (2005)]{matsumara05} 
 \textsc{Matsumura, S. \& Pudritz, R. E.} 2005, 
 \emph{ApJ}, 618, L137
 
\bibitem[Mayor \& Queloz (1995)]{mayor95}
 \textsc{Mayor, M. \& Queloz, D.} 1995, 
 \emph{Nature}, 378, 355
 
\bibitem[Menou \& Goodman (2004)]{menou04}
 \textsc{Menou, K. \& Goodman, J.} 2004, 
 \emph{ApJ}, 606,520 
 
\bibitem[Miyoshi et al. (1999)]{miyoshi99}	
 \textsc{Miyoshi, K., Takeuchi, T., Tanaka, H. \& Ida, S.} 1999, 
 \emph{ApJ}, 516, 451
 
\bibitem[Mizuno (1980)]{mizuno80} 	
 \textsc{Mizuno, H.} 1980,
 \emph{Progress of Theoretical Physics}, 64, 544

\bibitem[Murray et al. (1998)]{murray98} 
 \textsc{Murray, N., Hansen, B., Holman, M. \& Tremaine, S.} 1998, 
 \emph{Science}, 279, 69

\bibitem[Nelson \& Benz (2003a)]{nelson03a} 
 \textsc{Nelson, A. F. \& Benz, W.} 2003a, 
 \emph{ApJ}, 589, 578

\bibitem[Nelson \& Benz (2003b)]{nelson03b} 
 \textsc{Nelson, A. F. \& Benz, W.} 2003b, 
 \emph{ApJ}, 589, 556

\bibitem[Nelson \& Papaloizou (2003)]{nelson03}
 \textsc{Nelson, R. P. \& Papaloizou, J. C. B.} 2003, 
 \emph{MNRAS}, 339, 993
  
\bibitem[Nelson \& Papaloizou (2004)]{nelson04}
 \textsc{Nelson, R. P. \& Papaloizou, J. C. B.} 2004, 
 \emph{MNRAS}, 350, 849 

\bibitem[Nelson et al. (2000)]{nelson00} 
 \textsc{Nelson, R. P., Papaloizou, J. C. B., Masset, F. \& Kley, W.} 2000, 
 \emph{MNRAS}, 318, 18

\bibitem[Ogilvie (2001)]{ogilvie01} 
 \textsc{Ogilvie, G. I.} 2001, 
 \emph{MNRAS}, 325, 231 
 
\bibitem[Ogilvie \& Lubow (2003)]{ogilvie03} 
 \textsc{Ogilvie, G. I. \& Lubow, S. H.} 2003,
 \emph{ApJ}, 587, 398
 
\bibitem[Papaloizou (2002)]{papaloizou02}
 \textsc{Papaloizou, J. C. B.} 2002, 
 \emph{A\&A}, 388, 615 

\bibitem[Papaloizou \& Larwood (2000)]{papaloizou00} 
 \textsc{Papaloizou, J. C. B. \& Larwood, J. D.} 2000,
 \emph{MNRAS}, 315, 823 

\bibitem[Papaloizou \& Lin (1984)]{papaloizou84} 
 \textsc{Papaloizou, J. C. B. \& Lin, D. N. C.} 1984,
 \emph{ApJ}, 285, 818

\bibitem[Papaloizou, Nelson \& Masset (2001)]{papaloizou01b} 
 \textsc{Papaloizou, J. C. B., Nelson, R. P. \& Masset, F.} 2001, 
 \emph{A\&A}, 366, 263
 
\bibitem[Papaloizou, Nelson \& Snellgrove (2004)]{papaloizou04} 
 \textsc{Papaloizou, J. C. B., Nelson, R. P. \& Snellgrove, M. D.} 2004, 
 \emph{MNRAS}, 350, 829 

\bibitem[Papaloizou \& Terquem (1999)]{papaloizou99} 
 \textsc{Papaloizou, J. C. B. \& Terquem, C.} 1999, 
 \emph{ApJ}, 521, 823

\bibitem[Papaloizou \& Terquem (2001)]{papaloizou01} 
 \textsc{Papaloizou, J. C. B. \& Terquem, C.} 2001, 
 \emph{MNRAS}, 325, 221
 
\bibitem[Pollack et al. (1996)]{pollack96} 	
 \textsc{Pollack, J. B., Hubickyj, O., Bodenheimer, P., Lissauer, J. J., Podolak, M. \&
  Greenzweig, Y.} 1996, 
  \emph{Icarus}, 124, 62

\bibitem[Pringle (1991)]{pringle91}  
 \textsc{Pringle, J. E.} 1991,
 \emph{MNRAS}, 248, 754 

\bibitem[Rafikov (2002)]{rafikov02}  
 \textsc{Rafikov, R. R.} 2002, 
 \emph{ApJ}, 572, 566 
 
\bibitem[Rasio \& Ford (1996)]{rasio96} 	
 \textsc{Rasio, F. A. \& Ford, E. B.} 1996, 
 \emph{Science}, 274, 954

\bibitem[Rice \& Armitage (2003)]{rice03}
 \textsc{Rice, W. K. M. \& Armitage, P. J.} 2003, 
 \emph{ApJ}, 598, L55

\bibitem[Rice \& Armitage (2005)]{rice05}
 \textsc{Rice, W. K. M. \& Armitage, P. J.} 2005, 
 \emph{ApJ}, 630, 1107 
 
\bibitem[Salmeron \& Wardle (2005)]{salmeron05}
 \textsc{Salmeron, R.\& Wardle, M.} 2005, 
 \emph{MNRAS}, 361, 45
 
\bibitem[Sano et al. (2000)]{sano00}	
 \textsc{Sano, T., Miyama, S. M., Umebayashi, T. \& Nakano, T.} 2000, 
 \emph{ApJ}, 543, 486

\bibitem[Santos, Israelian \& Mayor (2004)]{santos04} 
 \textsc{Santos, N. C., Israelian, G. \& Mayor, M.} 2004, 
 \emph{A\&A}, 415, 1153
 
\bibitem[Sasselov \& Lecar (2000)]{sasselov00} 	
 \textsc{Sasselov, D. D. \& Lecar, M.} 2000, 
 \emph{ApJ}, 528, 995

\bibitem[Sch\"afer et al. (2004)]{schafer04} 
 \textsc{Sch\"afer, C., Speith, R., Hipp, M. \& Kley, W.} 2004, 
 \emph{A\&A}, 418, 325
 
\bibitem[Shakura \& Sunyaev (1973)]{shakura73} 
 \textsc{Shakura, N. I. \& Sunyaev, R. A.} 1973,
 \emph{A\&A}, 24, 337
 
\bibitem[Syer \& Clarke (1995)]{syer95}
 \textsc{Syer, D. \& Clarke, C. J.} 1995, 
 \emph{MNRAS}, 277, 758 
 
\bibitem[Takeuchi, Miyama \& Lin (1996)]{takeuchi96}	
 \textsc{Takeuchi, T., Miyama, S. M. \& Lin, D. N. C.} 1996, 
 \emph{ApJ}, 460, 832 
 
\bibitem[Tanaka \& Ida (1999)]{tanaka99}	
 \textsc{Tanaka, H. \& Ida, S.} 1999, 
 \emph{Icarus}, 139, 350 
 
\bibitem[Tanaka, Takeuchi \& Ward (2002)]{tanaka02}
 \textsc{Tanaka, H., Takeuchi, T. \& Ward, W. R.} 2002,
 \emph{ApJ}, 565, 1257 

\bibitem[Terquem (2003)]{terquem03} 
 \textsc{Terquem, C.} 2003, 
 \emph{MNRAS}, 341, 1157
 
\bibitem[Thommes (2005)]{thommes05} 	
 \textsc{Thommes, E. W.} 2005, 
 \emph{ApJ}, 626, 1033
 
\bibitem[Thommes, Duncan \& Levison (1999)]{thommes99} 
 \textsc{Thommes, E. W., Duncan, M. J. \& Levison, H. F.} 1999, 
 \emph{Nature}, 402, 635

\bibitem[Trilling et al. (1998)]{trilling98} 
 \textsc{Trilling, D. E., Benz, W., Guillot, T., Lunine, J. I., Hubbard, W. B. \& Burrows, A.} 1998, 
 \emph{ApJ}, 500, 428 
 
\bibitem[Trilling, Lunine \& Benz (2002)]{trilling02}	
 \textsc{Trilling, D. E., Lunine, J. I. \& Benz, W.} 2002, 
 \emph{A\&A}, 394, 241 
 
\bibitem[Tsiganis et al. (2005)]{tsiganis05} 
 \textsc{Tsiganis, K., Gomes, R., Morbidelli, A. \& Levison, H. F.} 2005, 
 \emph{Nature}, 435, 459
 
\bibitem[Veras \& Armitage (2004)]{veras04}
 \textsc{Veras, D. \& Armitage, P. J.} 2004, 
 \emph{MNRAS}, 347, 613 
 
\bibitem[Ward (1988)]{ward88} 
 \textsc{Ward, W. R.} 1988, 
 \emph{Icarus}, 73, 330 
 
\bibitem[Ward (1989)]{ward89} 	
 \textsc{Ward, W. R.} 1989,
 \emph{ApJ}, 345, L99
 
\bibitem[Ward (1997)]{ward97} 
 \textsc{Ward, W. R.} 1997, 
 \emph{Icarus}, 126, 261

\bibitem[Ward \& Hahn (1995)]{ward95} 
 \textsc{Ward, W. R. \& Hahn, J. M.} 1995, 
 \emph{ApJ}, 440, L25
 
\bibitem[Weidenschilling (1977)]{weidenschilling77}	
 \textsc{Weidenschilling, S. J.} 1977,
 \emph{Astrophysics and Space Science}, 51, 153 
 
\bibitem[Weidenschilling \& Marzari (1996)]{weidenschilling96} 
 \textsc{Weidenschilling, S. J. \& Marzari, F.} 1996, 
 \emph{Nature}, 384, 619

\bibitem[Winters, Balbus \& Hawley (2003a)]{winters03} 
 \textsc{Winters, W. F., Balbus, S. A. \& Hawley, J. F.} 2003a, 
 \emph{MNRAS}, 340, 519

\bibitem[Winters, Balbus \& Hawley (2003b)]{winters03b} 
 \textsc{Winters, W. F., Balbus, S. A. \& Hawley, J. F.} 2003b, 
 \emph{ApJ}, 589, 543

\end{thebibliography}
\end{document}